\documentclass[pdftex,11pt]{article}
\usepackage{amsfonts}
\usepackage{bm}
\usepackage{accents}
\usepackage[bulgarian,english]{babel}
\usepackage[T2A]{fontenc}

\begin{document}

\begin{flushright}
CERN-PH-TH/2011-050\\
\vspace{2.5mm}
To the memory of Matey Mateev,\\
a missing friend.
\end{flushright}
\begin{center}
{\Large {\bf Clifford Algebras and Spinors}\footnote{Lectures presented at the
University of Sofia in October-November, 2010. Lecture notes prepared together
with {\bf Dimitar Nedanovski} (e-mail: dnedanovski@inrne.bas.bg) and completed
during the stay of the author at the ICTP, Trieste and at CERN, Geneva.Published
 in {\it Bulg. J. Phys.} {\bf 38}:1 (2011) 3-28.}}

\vspace{2.5mm}

Ivan Todorov

Institute for Nuclear Research and Nuclear Energy\\
Tsarigradsko Chaussee  72, BG-1784 Sofia, Bulgaria\footnote{permanent address}\\
e-mail: todorov@inrne.bas.bg\\
and\\
Theory Group, Physics Department, CERN\\
CH-1211 Geneva 23, Switzerland
\end{center}
\begin{abstract}

Expository notes on Clifford algebras and spinors with a detailed discussion of
 Majorana, Weyl, and Dirac spinors. The paper is meant as a review of background material, needed, in particular, in now fashionable theoretical speculations on neutrino masses. It has a more mathematical flavour than the over twenty-seven-
year-old {\it Introduction to Majorana masses} \cite{M84} and includes
historical notes and biographical data on past participants in the story.

\end{abstract}

\tableofcontents

\newpage

\def\theequation{\thesection.\arabic{equation}}

\section{Quaternions, Grassmann and Clifford algebras}
\setcounter{equation}{0}
\renewcommand\theequation{\thesection.\arabic{equation}}

Clifford's\footnote{William Kingdon Clifford (1845-1879) early
appreciated the work of Lobachevsky and Riemann; he was the first to translate
into English Riemann's inaugural lecture {\it On the hypotheses which lie at
the bases of geometry}. His view of the physical world as {\it variation of
curvature of space} anticipated Einstein's general theory of relativity. He
died (before reaching 34) of tuberculosis, aggravated (if not caused) by
overwork.} paper \cite{Cl} on ``geometric algebra'' (published a year before
his death) had two sources: Grassmann's \footnote{Hermann G\"unter Grassmann
(1809-1877), a German polymath, first published his fundamental work that led
the foundations of linear algebra (and contained the definition of {\it
exterior product}), in 1844. He was too far ahead of his time to be understood
by his contemporaries. Unable to get a position as a professor in mathematics,
Grassmann turned to linguistic. His sound law of Indo-European (in particular,
of Greek and Sanskrit) languages was recognized during his lifetime.} algebra
and Hamilton's \footnote{William Rowan Hamilton (1805-1865) introduced during
1827-1835 what is now called {\it Hamiltonian} but also the Lagrangian formalism
 unifying mechanics and optics. He realized by that time that multiplication by
 a complex number of absolute value one is equivalent to a rotation in the
euclidean (complex) 2-plane ${\mathbb C}$ and started looking for a
3-dimensional generalization of the complex numbers that would play a similar
role in the geometry of 3-space. After many unsuccessful attempts, on October
16, 1843, while walking along the Royal Canal, he suddenly had the inspiration
that not three but a four dimensional generalization of ${\mathbb C}$ existed
and was doing the job -- see introduction to \cite{B}.} quaternions whose three
 imaginary units $i, j, k$ can be characterized by
\begin{equation}
 i^2 = j^2 = k^2 = ijk = -1.
\label{eq1.1}\end{equation}
We leave it to the reader to verify that these equations imply $ij = k = -ji,
jk = i = - kj, ki = j = -ik$.

We proceed to the definition of a (real) Clifford algebra and will then
display the Grassmann and the quaternion algebras as special cases.

Let $V$ be a real vector space equipped with a quadratic form $Q(v)$ which
gives rise - via {\it polarization} - to a symmetric bilinear form $B$ such that
 $2B(u,v) = Q(u+v) - Q(u) - Q(v)$. The {\it Clifford algebra} $Cl(V, Q)$ is
the associative algebra freely generated by $V$ modulo the relations
\begin{equation}
v^2 = Q(v) (= B(v, v))\ {\rm for\ all}\ v \in V\,,\ \Leftrightarrow uv + vu =
2B(u, v) \equiv 2(u, v).
\end{equation}
(Here and in what follows we identify the vector $v \in V$ with its image, say,
 $i(v)$ in $Cl(V, Q)$ and omit the symbol $1$ for the algebra unit on the right
 hand side.) In the special case $B=0$ this is the {\it exterior} or {\it
Grassmann} algebra $\Lambda(V)$, the direct sum of skewsymmetric tensor
products of $V = {\mathbb R}^n$:
\begin{equation}
\Lambda (V) = \oplus_{k=0}^n \Lambda^k (V) \ \Rightarrow \
dim\Lambda (V) = \sum_{k=0}^n \left({n}\atop{k}\right) = (1 + 1)^n = 2^n.
\label{eq1.3}
\end{equation} Having in mind applications to
the algebra of $\gamma$-matrices we shall be interested in the
opposite case in which $B$ is a non-degenerate, in general
indefinite, real symmetric form:
\begin{equation}
Q(v) = (v, v) = v_1^2 + ... + v_p^2 - v_{p+1}^2 - ... - v_n^2\,,\quad n = p+q.
\label{eq1.4}
\end{equation}
We shall then write $Cl(V, Q) = Cl(p,q)$, using the shorthand notation $Cl(n,0)
 = Cl(n),\ Cl(0,n) =Cl(-n)$ in the euclidean (positive or negative definite)
case.\footnote{Mathematicians often use the opposite sign convention
corresponding to $Cl(n) = Cl(0,n)$ that fits the case of normed star algebras
-- see \cite{B} which contains, in particular, a succinct survey of
Clifford algebras in Sect. 2.3. The textbook \cite{L} and the (46-page-long,
mathematical) tutorial on the subject \cite{G08} use the same sign convention
as ours but opposite to the  monograph \cite{LM}. The last two references rely
on the modern classic on Clifford modules \cite{ABS}.} The expansion
(\ref{eq1.3}) is applicable to an arbitrary Clifford algebra providing a
{\it ${\mathbb Z}$ grading} for any $Cl(V) \equiv Cl(V, Q)$ {\it as a vector
space} (not as an algebra). To see this we start with a basis $e_1, ..., e_n$ of
 orthogonal vectors of $V$  and define a linear basis of $Cl(V)$ by the sequence
 \begin{eqnarray}
&& 1, ..., (e_{i_1}...e_{i_k}, \, 1 \leq i_1 < i_2 < ... < i_k \leq n), \,
k = 1, 2, ..., n  \,                           \nonumber        \\
&& (2 e_ie_j = [e_i, e_j] \, {\rm for} \, i<j).
\label{basis}
\end{eqnarray}
It follows that the dimension of $Cl(p, q)$ is again $2^n (n=p+q)$. We leave it
 as an exercise to the reader to prove that $Cl(0)={\mathbb R},\ Cl(-1) =
{\mathbb C},\ Cl(-2) = {\mathbb H}$ where ${\mathbb H}$ is the algebra of
quaternions; $Cl(-3) = {\mathbb H} \oplus {\mathbb H}$. ({\it Hint}: if $e_\nu$
 form an orthonormal basis in $V$ (so that $e_\nu^2 = -1$) then in the third
case, set $e_1 = i, e_2 = j, e_1 e_2 = k$ and verify the basic relations
(\ref{eq1.1}); verify that in the fourth case the operators
${}^1/_2(1\pm e_1 e_2 e_3)$
 play the role of orthogonal projectors to the two copies of the quaternions.)
An instructive example of the opposite type is provided by the algebra $Cl(2)$.
 If we represent in this case the basic vectors by the real $2\times 2$ Pauli
matrices: $e_1 = \sigma_1, e_2 = \sigma_3$ we find that $Cl(2)$ is isomorphic to
${\mathbb R}[2]$, the algebra of all real $2\times 2$ matrices. If instead we
set $e_2 = \sigma_2$ we shall have another algebra (over the real numbers) of
complex $2\times 2$ matrices. An invariant way to characterize $Cl(2)$ (which
embraces the above two realizations) is to say that it is isomorphic to the
complex $2\times 2$ matrices invariant under an ${\mathbb R}$-linear involution
 given by the complex conjugation K composed with an inner automorphism. In the
 first case the involution is just the complex conjugation; in the second it is
 $K$ combined with a similarity transformation: $x \rightarrow \sigma_1 K x
 \sigma_1$.

We note that $Cl(-n), n = 0, 1, 2$ are the only {\it division rings} among the Clifford algebras. All others have {\it zero divisors}. For instance, $(1 + e_1 e_2 e_3)(1 - e_1 e_2 e_3) = 0$ in $Cl(-3)$ albeit none of the two factors is zero.

Clifford algebras are {\it ${\mathbb Z}_2$ graded}, thus providing
an example of {\it superalgebras}. Indeed, the linear map $v \rightarrow -v$ on
 $V$ which preserves $Q(v)$ gives rise to an involutive automorphism $\alpha$
of $Cl(V, Q)$. As $\alpha^2 = id$ (the identity automorphism) - the defining
property of an involution - it has two eigenvalues, $\pm 1$; hence $Cl(V)$
splits into a direct sum of {\it even} and {\it odd} elements:
\begin{equation}
Cl(V) = Cl^0(V) \oplus Cl^1(V), Cl^i(V) =
\oplus_{k=0}^{[n/2]}\Lambda^{i+2k}V,  \,  i = 0, 1.
\label{Z2}
\end{equation}

{\it Exercise 1.1} Demonstrate that $Cl^0(V, Q)$ is a Clifford subalgebra of
$Cl(V, Q)$; more precisely, prove that if $V$ is the orthogonal direct sum of a
1-dimensional subspace of vectors collinear with $v$ and a subspace $U$ then
$Cl^0(V, Q) = Cl(U, -Q(v)Q|_U)$ where $Q|_U$ stands for restriction of the form
 $Q$ to $U$. Deduce that, in particular,
\begin{equation}
Cl^0(p, q)\simeq Cl(p, q-1)\ \ {\rm for}\ q>0\,,\quad Cl^0(p, q)\simeq
Cl(q, p-1)\ \ {\rm for}\ p>0\,.
\label{Cl0}
\end{equation}
In particular, for the algebra $Cl(3, 1)$ of Dirac \footnote{Paul Dirac
(1902-1984) discovered his equation (the ``square root'' of the d'Alembertian)
describing the electron and predicting the positron in 1928
\cite{D28}. At 31, in 1933, he was the youngest theoretician ever to
 be awarded the Nobel Prize in physics. His quiet life and strange character
are featured in the widely acclaimed biography \cite{F}.} $\gamma$-matrices the
 even subalgebra (which contains the generators of the Lorentz Lie algebra) is
isomorphic to $Cl(3) \simeq Cl(1, 2)$ (isomorphic as algebras, not as
superalgebras: their gradings are inequivalent).

We shall reproduce without proofs the classification of real Clifford algebras.
(The examples of interest will be treated in detail later on.) If $R$ is a ring,
we denote by $R[n]$ the algebra of $n\times n$ matrices with entries in $R$.

{\bf Proposition 1.1} {\it The following symmetry relations hold:}
\begin{equation}
Cl(p+1, q+1) = Cl(p, q)[2],\ Cl(p+4, q) = Cl(p, q+4).
\label{ClSym}
\end{equation}
{\it They imply the Cartan-Bott\footnote{{\it \'Elie Cartan} (1869-1951)
developed the theory of Lie groups and of (antisymmetric) differential forms.
He discovered the 'period' 8 in 1908 - see \cite{B} \cite{CCh} where the
original papers are cited. {\it Raoul Bott} (1923-2005) established his version
of the periodicity theorem in the context of homotopy theory of Lie groups in
1956 - see \cite{Tu} and references therein.} periodicity theorem}
\begin{equation}
Cl(p+8, q) = Cl(p+4, q+4) = Cl(p, q+8) = Cl(p, q)[16] = Cl(p, q)\otimes
{\mathbb R}[16].
\label{mod8}
\end{equation}

Let $(e_1, ..., e_p, e_{p+1}, ..., e_n), n = p+q$ be an orthonormal basis in
$V$, so that
\begin{equation}
(e_i, e_j) (= B(e_i, e_j)) = \eta_{ij}:= e_i^2 \delta_{ij, },\
e_1^2 = ... = e_p^2 = ... = -e_n^2 = 1.
\label{eta}
\end{equation}
Define the {\it (pseudoscalar) Coxeter\footnote{(H.S.M.) Donald Coxeter
(1907-2003) was born in London, but worked for 60 years at the University of
Toronto in Canada. An accomplished pianist, he felt that mathematics and music
were intimately related. He studied the product of reflections in 1951.}
``volume'' element}
\begin{equation}
\omega = e_1 e_2 ... e_n\ \Rightarrow\ \omega^2 = (-1)^{(p-q)(p-q-1)/2}.
\label{omega}\end{equation}

{\bf Proposition 1.2} {\it The types of algebra $Cl(p, q)$ depend on
$p-q\:\:\rm {mod}\,8$ as displayed on Table 1:}
\begin{table}[!hbp]
\begin{tabular}{| c | c | c | c | c | c |}
\hline
$p-q\:\:\rm{mod}\,8$ & $\omega^2$ & \textit{Cl}(p,q) & $p-q\:\:\rm{mod}\,8$   & $\omega^2$ & \textit{Cl}(p,q)\\
               &            &$p+q=2m$ &                  &           & $p+q=2m+1$\\
\hline
$0$            & $+$        &$\mathbb{R}$$[2^m]$ & $1$   & $+$   & $\mathbb{R}$$[2^m]\oplus\mathbb{R}$$[2^m]$\\
\hline
$2$            & $-$       &$\mathbb{R}$$[2^m]$  & $3$   & $-$   &$\mathbb{C}$$[2^m]$\\
\hline
$4$            & $+$       &$\mathbb{H}$$[2^{m-1}]$ & $5$  & $+$ &
$\mathbb{H}$$[2^{m-1}]\oplus\mathbb{H}$$[2^{m-1}]$\\
\hline
$6$            & $-$       &$\mathbb{H}$$[2^{m-1}]$ & $7$  & $-$ &$\mathbb{C}$$[2^m]$\\
\hline
\end{tabular}
\caption{}
\end{table}\\

The reader should note the appearance of a complex matrix algebra in two of the
 above realizations of $Cl(p, q)$ for odd dimensional real vector spaces. The
algebra $Cl(4, 1) = {\mathbb C}[4] (= Cl(2,3))$ is of particular interest: it
appears as an extension of the Lorentz Clifford algebra $Cl(3, 1)$ (as well as
of $Cl(1, 3)$). As we shall see later (see Proposition 2.2, below) $Cl(4, 1)$ gives rise in a natural way to the central extension $U(2, 2)$ of the spinorial
conformal group and of its Lie algebra.

{\it Exercise 1.2} Prove that for $n (= p+q)$ odd the Coxeter element  of the
algebra $Cl(p, q)$ is central and defines a complex structure for $p - q = 3$ mod 4. For $n$ even its ${\mathbb Z}_2$-graded
commutator with homogeneous elements vanish:
\begin{equation}
\omega x_j = (-1)^{j(n-1)} x_j \omega \, \, {\rm for} j = 0, 1.
\label{eq1.12}
\end{equation}

For proofs and more details on the classification of Clifford algebras - see
\cite{L}, Sect. 16, or \cite{LM} (Chapter I, Sect. 4) where also a better
digested ``Clifford chessboard'' can be found (on p. 29). The classification
for $q = 0, 1$ can be extracted from the matrix representation of the Clifford
units, given in Sect. 3.

{\bf Historical note}. The work of Hamilton on quaternions was appreciated and continued by Arthur Cayley (1821-1895), "the greatest English mathematician of the last century - and this", in the words of H.W. Turnbull (of 1923) \cite{Cr}. Cayley rediscovered (after J.T. Graves) the octonions in 1845. Inspired and supported by Cayley in his student years, Clifford defined his {\it geometric algebra} \cite{Cl} (discovered in 1876) as generated by $n$ orthogonal unit vectors, ${\bf e}_1, ..., {\bf e}_n$, which anticommute, ${\bf e}_i{\bf e}_j = -{\bf e}_j{\bf e}_i$ (like in Grassmann) and satisfy ${\bf e}_i^2 = -1$ (like in Hamilton), both preceding papers appearing in 1844 (on the eve of Clifford's birth). In a subsequent article, published posthumously, in 1882, Clifford also considered the algebra $Cl(n)$ with ${\bf e}_i^2 = 1$ for all $i$. He distinguished four classes of geometric algebras according to two sign factors: the square of the
Coxeter element (\ref{omega}) and the factor $(-1)^{n-1}$ appearing in $\omega
e_i = (-1)^{n-1} e_i \omega$ (cf. (\ref{eq1.12})). It was \'Elie Cartan who
identified in 1908 the general Clifford algebras $Cl(p, q)$ with matrix
algebras with entries in ${\mathbb R}, {\mathbb C}, {\mathbb H}$ and found the
period $8$ as displayed in Table 1. A nostalgic survey of quaterions and their
possible applications to physics is contained in the popular article \cite{La}.
 A lively historical account of Clifford algebras and spinors is given by
Andrzej Trautman - see, in particular, the first reference \cite{Tr} as well as
 in his book \cite{BT}, written jointly with Paolo Budinich - the physicist who
was instrumental in founding both the ICTP and the International School for
Advanced Studies (SISSA-ISAS) in Trieste, and is a great enthusiast of Cartan
spinors.

\newpage

\section{The groups $Pin(p, q)$ and $Spin(p, q)$; conjugation and norm}
\setcounter{equation}{0}
\renewcommand\theequation{\thesection.\arabic{equation}}

Define the unique antihomomorphism $x \rightarrow x^\dagger$ of $Cl(V)$ called
{\it conjugation} for which
\begin{equation}
v^\dagger = - v\ {\rm for\ all}\ v \in V\ ({\rm and}\ (xy)^\dagger = y^\dagger
x^\dagger \, {\rm for}\  x, y \in Cl(V)).
\label{bar}
\end{equation}
Whenever we consider a complexification of our Clifford algebra we will extend
this antihomomorphism to an antilinear antiinvolution (that is, we assume that
$(cx)^\dagger = {\bar c} x^\dagger$ for any $c \in {\mathbb C}, x \in Cl(V)$,
where the bar stands for complex conjugation). We shall say that an element
$x \in Cl(V)$ is {\it pseudo\-(anti)\-hermitean} if $x^\dagger = (-)x$.
The notion of conjugation allows to define a map
\begin{equation}\label{2.2}
   N: Cl(V) \longrightarrow Cl(V),\; N(x)= x x^\dagger(=x^\dagger x),
\end{equation}
called {\it norm}. It extends, in a sense, the quadratic form $-Q$ to the full
Clifford algebra and coincides with the usual /positive/ ``norm squared'' on
the quaternions:
\[
N(s + x{\bf i} + y{\bf j} + z{\bf ij} )= s^2 + x^2 + y^2 + z^2
\, {\rm for} \ s + x{\bf i} + y{\bf j} + z{\bf ij} \in Cl(-2).
\]
For products of vectors of $V, N(x)$ is a scalar: one easily verifies the
implication
\begin{equation}
x = v_1 ... v_k \Rightarrow x x^\dagger = (-1)^k Q(v_1) ... Q(v_k)
(= N(v_1...v_k)).
\label{NQ}
\end{equation}
This would suffice to define the groups $Pin(n)$ and $Spin(n)$ as products of
Clifford units (cf. Sect. 2.4 of \cite{B}). We shall sketch here the more
general approach of \cite{ABS} and \cite{LM}(digested in the ``tutorial on
Clifford algebra and the groups {\bf Spin} and {\bf Pin}'' \cite{G08}).

Let $Cl(p, q)^*$ be the group of invertible elements of $Cl(p, q)$. It seems
natural to use its {\it adjoint action} on $V, {\mathrm{Ad_x}} v: = xvx^{-1}$, to define a
covering of the ortrhogonal group $O(p, q)$ as it automatically preserves the
quadratic form (\ref{eq1.4}): $(xvx^{-1})^2 = v^2$ (provided $x\in Cl(p,q)^*$ is
such that ${\mathrm{Ad_x}} v\in V$ for all $v\in V$). The adjoint action, however,
does not contain the reflections
\begin{equation}
-u v u^{-1} = v - 2\frac{(u, v) u}{u^2}, \, {\rm for} \, u \in V, \,u^2\neq0,\, u^{-1} =
u/u^2,
\label{refl}
\end{equation}
for an odd dimensional $V$. To amend this we shall use, following \cite{ABS}
and \cite{LM}, a {\it twisted adjoint} representation. We define the {\it
Clifford} (or {\it Lipschitz\footnote{The German mathematician Rudolf Lipschitz
(1832-1903) discovered independently the Clifford algebras in 1880 and
introduced the groups $\Gamma_{0, n}$ - see the appendix {\it A history of
Clifford algebras} in \cite{L}}) group} $\Gamma_{p.q}$ through its action on
$V = {\mathbb R}^{p,q}$:
\begin{equation}
x \in \Gamma_{p,q} \, {\rm iff} \, \rho_x: v \rightarrow \alpha(x) v x^{-1} \in
 V, \ {\rm for \, any} \, v \in V,
\label{Clgr}
\end{equation}
where $\alpha$ is the involutive automorphism which maps each odd element
$x \in Cl^1(V)$ (in particular, each element in $V$) to $-x$ (the involution
$\alpha$ was, in fact, used in Sect. 1 to define the ${\mathbb Z}_2$-grading on
 $Cl(p, q)$). It is not obvious that the map (\ref{Clgr}) preserves the
 form $Q(v) = v^2$ (\ref{eq1.4}) since $\alpha(x)\neq\pm x$, for inhomogeneous
$x \in \Gamma_{p,q}$ .The following theorem verifies it and gives a more precise picture.

{\bf Theorem 2.1} {\it The map $\rho: \Gamma_{p,q} \rightarrow
O(p,q)$ is a surjective homomorphism whose kernel is the multiplicative group ${\mathbb R}^* 1$ of the nonzero scalar multiples of the Clifford unit. The restriction of $N(x)$ to $\Gamma_{p,q}$ is a nonzero scalar.}

In other words, every element (including reflections) of $O(p, q)$ is the image (under (\ref{Clgr})) of some element $x \in \Gamma_{p,q}$, and, furthermore, if $x$ satisfies $\alpha(x) v = v x$ for all $v \in V$, then $x$ is a real number (times the Clifford unit).

\smallskip
In order to prove the last statement, we separate the even and the odd part of $x$: $x=x_0+x_1$, $\alpha(x)=x_0-x_1$. Assuming that
\begin{equation}\label{eq1_proof_Th.2.1}
\alpha(x)v\big(\!\!=(x_0-x_1)v\big)=vx\big(\!\!=v(x_0+x_1)\big)\textrm{ for all } v\in V,
\end{equation}
we shall prove that $x_1=0$ while $x_0\in\mathbb{R}^*$. To this end we expand $x_0$ and $x_1$ in the basis (\ref{basis}). We shall prove that neither expansion contains the vector $e_n$. Let indeed $x_0=a_0+a_1e_n$, $x_1=b_0e_n+b_1$ where $a_0$ and $b_0$ are even while $a_1$ and $b_1$ are odd elements of the Clifford algebra, independent of $e_n$. According to our assumption
\begin{eqnarray}
    (a_0+a_1e_n)v=v(a_0+a_1e_n)\textrm{ for all } v\in V.\nonumber
\end{eqnarray}
Taking $v=e_n$ we find $a_1e_n^2=e_na_1e_n=-a_1e_n^2$, as $e^2_n=\pm1$, we conclude that $a_1=0$. Similarly, $-(b_0e_n+b_1)v=v(b_0e_n+b_1)$ implies,  for $v=e_n$, that $-b_0e_n^2=b_0e_n^2$, i.e. $b_0=0$. It follows that $x_0$ and $x_1$, and hence $x$, are independent of $e_n$. A similar argument is valid for any of the basic vectors $e_i$.
Consequently (\ref{eq1_proof_Th.2.1} implies that $x$ is a real multiple of the unit element of $Cl(V)$.
\par On the other hand, the implication
\begin{eqnarray}
\alpha(x)vx^{-1}=u\in V\, =>\, \alpha(x)vx^{-1}=\alpha(x^{\dagger-1})vx^{\dagger}\nonumber
\end{eqnarray}
(obtained by noting that $\alpha(u^{\dagger})=u$ for any $u\in V$) yields the relation
\begin{equation}\label{eq2_proof_Th.2.1}
    \alpha\big(N(x)^{-1}\big)vN(x)=v\textrm{ (if } \alpha(x)vx^{-1}\in V)
\end{equation}
Therefore, $x\in \Gamma_{p,q}$ indeed implies $N(x)\in\mathbb{R}^*.$
\par The above proof of Theorem 2.1 ia an adaptation of the argument of
\cite{LM} (Chapter I, Sect. 2)- see also \cite{G08} (Lemma 1.7 and Proposition 1.8).
\medskip
\par {\it Exercise 2.1} Let $(e_0,e_1)$ be an orthonormal basis of $Cl(1,1)$ $(e_1^2=1=-e_0^2,\, e_0e_1+e_1e_0=0)$. Prove that the rotated basis
\begin{eqnarray*}
   && v_0(\beta)=\cosh\beta e_0+\sinh\beta e_1 \\
   && v_1(\beta)=\sinh\beta e_0+\cosh\beta e_1
\end{eqnarray*}
can be represented as a superposition of two reflections. (Hint: use the relations
\begin{eqnarray*}
  &&-v_1(\frac{\beta}{2})e_0v_1(\frac{\beta}{2})=v_0(\beta),\,\textrm{ }(v_0^2(\beta)=-1) \\
  &&-v_1(\frac{\beta}{2})e_1v_1(\frac{\beta}{2})=-v_1(\beta),\,\textrm{ }(v_1^2(\beta)=1)
\end{eqnarray*}
and the identities $-v_1v_0v_1=v_0,\,-v_1v_1v_1=-v_1$ that use the anticommutativity of $v_0$ and $v_1$.)


The group {\it $Pin(p, q)$} is defined as the subgroup of $\Gamma_{p,q}$ of
elements $x$ for which $N(x) = \pm 1$. The restriction of the map $\rho$ to {\it $Pin(p, q)$}
gives rise to a (two-to-one) homomorphism of $Pin(p, q)$ on the orthogonal group $O(p, q)$.
\par The proof of this statement for the compact case (for which $pq=0$) is standard- see e.g. Theorem 2.7 (Cartan-Dieudonn\'{e} theorem) of \cite{LM} (p.17).
 The group $Spin(p, q)$ is obtained as the intersection of $Pin(p, q)$
with the even subalgebra $Cl^0(p, q)$.\smallskip

For any vector $v$ in $V \subset Cl(p, q)$
each element $x$ of $Spin(p, q)$ defines a map preserving $Q(v)$
(we note that for $x \in Spin(p, q), \alpha(x) =x$,
so that the twisted adjoint coincides with the standard one):
\begin{equation}
v \rightarrow xvx^{-1}\quad (x^{-1} = N(x) x^\dagger, \, {\rm for} N(x)^2 = 1).
\label{adj}
\end{equation}
The (connected) group $Spin(p, q)$ can
be defined as the double cover of the identity component $SO_0(p, q)$ of
$SO(p, q)$ and is mapped onto it under (\ref{adj}). The Lie algebra
$spin(p, q)$ of the Lie group $Spin(p, q)$ is generated by the
commutators $[e_i, e_j]$ of a basis of $V = {\mathbb R}^{(p,q)}$.

{\it Remark 2.1} Another way to approach the spin groups starts with the
observation that the (connected) orthogonal group $SO(n)$ is not simply
connected, its {\it fundamental (or homotopy) group\footnote{Anticipated by
Bernhard Riemann (1826-1866), the notion of fundamental group was introduced by
 Henri Poincar\'e (1854-1912)in his article {\it Analysis Situs} in 1895.}}
consists of two elements, $\pi_1(SO(n)) = {\mathbb Z}_2$ for $n>2$, while for
the circle, $n = 2$, it is infinite: $\pi_1(SO(2) = {\mathbb Z}$. The homotopy
group of the pseudo-orthogonal group $SO(p, q)$ is equal to that of its maximal
 compact subgroup:
\begin{equation}
\pi_1(SO_0(p, q)) = \pi_1(SO(p)) \times \pi_1(SO(q))(= {\mathbb Z}_2 \, {\rm
for}\ \, p > 2, \, q \leq 1).
\label{hom}
\end{equation}
In all cases the group $Spin(p, q)$ can be defined as the double cover of
$SO_0(p, q)$ (which coincides with its universal cover for $p > 2, q \leq 1$).

{\it Exercise 2.2}\ Verify that the Coxeter element $\omega$ (\ref{omega})
generates the centre of $Spin(p, q)$ for $p-q \neq 4$ mod 8 while the centre of
$Spin(4\ell)$ is ${\mathbb Z}_2 \times {\mathbb Z}_2$ (see Appendix A1 to
\cite{KT}).

\smallskip

We proceed to describe the {\it spinor representations}\footnote{The theory of
finite dimensional irreducible representations of (semi)simple Lie groups
(including the spinors) was founded by E. Cartan in 1913 - see the historical
survey\cite{CCh}. The word {\it spinor} was introduced by Paul Ehrenfest
(1880-1933) who asked in the fall of 1928 the Dutch mathematician B.L. van der
Waerden (1903-1996) to help clear up what he called the ``group plague'' (see
\cite{Sch} and Lecture 7 in \cite{To}).} in low dimensions. More precisely, we
shall identify $spin(p, q)$ and $Spin (p, q)$ as a sub-Lie-algebra and a
subgroup in $Cl(p, q)$. As it is clear from Table 1 for $n(= p + q)= 2m$ there
is a single irreducible {\it Clifford module} of dimension $2^m$; for $n = 2m +
 1$ there may be two irreducible representations of the same dimension. In
either case, knowing the embedding of the spin group into the Clifford algebra
we can thereby find its defining representation.

Consider the 8-dimensional Clifford algebra $Cl(3)$ spanned by the unit scalar,
 $1$, the three orthogonal unit vectors, $\sigma_j, j= 1, 2, 3$, the unit
bivectors $\sigma_1 \sigma_2, \sigma_2 \sigma_3, \sigma_3 \sigma_1$, and the
pseudoscalar $(\omega_3 \equiv) i:= \sigma_1 \sigma_2 \sigma_3$. It is
straightforward to show that the conditions $\sigma_j^2 = 1$ and the
anticommutativity of $\sigma_j$ imply
\begin{equation}
(\sigma_1 \sigma_2)^2 = (\sigma_2 \sigma_3)^2 = (\sigma_3 \sigma_1)^2 = -1 = i^2.
\label{sigma}
\end{equation}
(The $\sigma_j$ here are just the unit vectors in ${\mathbb R}^3$ that generate
$Cl(3)$. We do not use the properties of the Pauli matrices which can serve as
their representation.) The subalgebra $Cl^0(3)$ spans the 4-dimensional space
$Cl(-2) = {\mathbb H}$ of quaternions, thus illustrating the relation
(\ref{Cl0}). It contains a group of unitaries of the form
\begin{eqnarray}
U \!\!\!&=&\!\!\! cos(\theta/2) - (n_1\sigma_2 \sigma_3 + n_2\sigma_3 \sigma_1 + n_3\sigma_1
\sigma_2) sin(\theta/2) = \nonumber\\
\!\!\!&=&\!\!\! cos(\theta/2) - i{\bf n \boldsymbol{\sigma}}sin(\theta/2), \ {\bf n}^2 = 1 , \
{\bf n\boldsymbol{ \sigma}} = n_1 \sigma_1 + n_2 \sigma_2 + n_3 \sigma_3,
\label{SU}
\end{eqnarray}
that is isomorphic to $SU(2)$. Furthermore, the transformation of 3-vectors $v$
given by (\ref{adj}) with $U^{-1} = U^* (= U^\dagger)$ where $\sigma_j^* =
\sigma_j, i^*=-i$ represents an $SO(3)$ rotation on angle $\theta$ around the
axis ${\bf n}$. The map $SU(2) \rightarrow SO(3)$ thus defined is two-to-one as
 $U =-1$ corresponds to the identity $SO(3)$ transformation.

The 16-dimensional euclidean algebra $Cl(4)$ generated by the orthonormal vectors
$\gamma_\alpha$ such that $[\gamma_\alpha, \gamma_\beta]_+ =2 \delta_{\alpha \beta}, \, \alpha, \beta = 1, 2, 3, 4$ is isomorphic to ${\mathbb H}[2]$. Its even part is given by the algebra $Cl(-3)$ discussed in Sect. 1: $Cl^0(4)\simeq Cl(-3) \simeq {\mathbb H} \oplus {\mathbb H}$. The corresponding Lorentzian\footnote{Hendrik Antoon Lorentz (1853-1928) introduced his transformations describing electromagnetic phenomena in the 1890's. He was awarded the Nobel Prize (together with his student Pieter Zeeman (1865-1943)) ``for their research into the influence of magnetism upon radiation phenomena''.} Clifford algebra $Cl(3, 1)$ is generated by the orthonormal elements $\gamma_\mu$ satisfying
\begin{equation}
[\gamma_\mu, \gamma_\nu]_+ = 2\eta_{\mu\nu}, \, \mu, \nu = 0, 1, 2, 3, \,
(\eta_{\mu\nu}) = diag(-1, 1, 1, 1).
\label{gamma31}
\end{equation}

According to (\ref{Cl0}) the even subalgebra $Cl^0(4, 1)$ is isomorphic to
the above $Cl(4)\simeq {\mathbb H}[2]$ while $Cl^0(3, 1) \simeq Cl(3) \simeq {\mathbb C}[2]$. It contains both the generators $\gamma_{\mu\nu}:= 1/2[\gamma_\mu, \gamma_\nu]$ of the Lie algebra $spin(3,1)$ and the elements of the spinorial Lorentz group $SL(2,{\mathbb C})$. It is easy to verify that  the elements $\gamma_0 \gamma_j$ (corresponding to $\sigma_j$ in $Cl(3)$) have square one while the pseudoscalar (\ref{omega}) $\omega (=
\omega_{3,1}) = \gamma_0 \gamma_1 \gamma_2 \gamma_3$ satisfies $\omega^2 = -1$ and
\begin{equation}
\gamma_{12} = \omega \gamma_{03}, \, \gamma_{31} =\omega \gamma_{02}, \,
\gamma_{23} = \omega \gamma_{01}.
\label{omgam}
\end{equation}
It follows that every even element of $Cl(3, 1)$ can be written in the form
\begin{equation}
z = z^0 + z^j \gamma_{0j}, \, z^\mu = x^\mu + \omega y^\mu, \, \mu =0, ...,3, \,
x^\mu, y^\mu \in {\mathbb R},
\label{evenCl}
\end{equation}
thus displaying the complex structure generated by the central element $\omega$ of
$Cl^0(3, 1)$ (of square $-1$). In particular, the Lie algebra $spin(3,1)$ generated by $z^j \gamma_{0j}$ is nothing but $sl(2, {\mathbb C})$. The group $Spin(3, 1)$ (a special case of $Spin(p, q)$ defined in the beginning of this section) is isomorphic to $SL(2, {\mathbb C})$, the group of complex $2 \times 2$ matrices of determinant one (which appears as the simply connected group with the above
Lie algebra).

{\bf Proposition 2.2 (a)} {\it The pseudoantihermitean elements $x\in Cl(4, 1)$ (satisfying $x^\dagger = -x$)  span the 16-dimensional Lie algebra $u(2,2)$. The corresponding Lie group $U(2, 2)$ consists of all pseudounitary elements $u \in
Cl(4, 1), u u^\dagger = 1$. There exists a (unique up to normalization) $U(2, 2)$-invariant sesquilinear form
$\widetilde{\psi} \psi = \psi^*\beta \psi$ in the space ${\mathbb C}^4$  of 4-component spinors (viewd as a
$Cl(4, 1)$-module) where the element $\beta$ of  $Cl(4, 1)$ intertwines the standard hermitean conjugation $*$
of matrices with the antiinvolution (\ref{bar}):}
\begin{eqnarray}
&& \gamma_a^* \beta = - \beta \gamma_a, \Rightarrow \gamma_{a b}^* \beta =
- \beta \gamma_{a b}, \, a, b = 0, 1, 2, 3, 4; \, \Rightarrow x^* \beta = \beta x^\dagger; \nonumber \\
&& u^* \beta = \beta u^{-1} \, \,{\rm for} \, u \in U(2, 2).
\label{beta}
\end{eqnarray}
{\bf (b)} {\it The intersection of $U(2,2)$ with $Cl(3, 1)$
coincides with the 10-parameter real symplectic group $Sp(4, {\mathbb R})\simeq
Spin(3, 2)$ whose Lie algebra $sp(4, {\mathbb R})$ is spanned by $\gamma_{\mu\nu}$ and by the odd elements $\gamma_\mu \in Cl^1(3, 1)$. The corresponding symplectic form is expressed in terms of the charge conjugation matrix $C$, defined in Sect.
3 below. An element $\Lambda = c_0 + \sum_{j=1}^3 c_j \gamma_{0j}$ of
$Cl^0(3, 1)$, $c_\nu = a_\nu + \omega b_\nu, \,{a_\nu, b_\nu} \in {\mathbb R}$
belongs to $Spin(3, 1) \subset Spin(3, 2)$ iff $N(\Lambda) = c_0^2 - {\bf c}^2 = 1$ where ${\bf c}^2 = \sum_{i=1}^3 c_i^2$.}                    \\

We leave the proof to the reader, only indicating that $u(2, 2)$ is spanned by
$\gamma_a, \gamma_{ab}$, and by the central element $\omega_{4,1}$ which plays the role of the imaginary unit.

{\it Exercise 2.3} Verify that space and time reflections are given by the odd
elements
\begin{equation}
\Lambda_s = \gamma^0 \, (\Lambda_s^{-1} = \gamma_0 = -\gamma^0), \,
\Lambda_t = \gamma^0 \omega \, \, (\Lambda_t^{-1} = \gamma_0\omega).
\label{st}
\end{equation}
Prove that, in general, for $\Lambda \in Pin(3, 1)$,
\begin{equation}
\Lambda \gamma p \Lambda^{-1} = \gamma L(\Lambda)p, \,p\gamma:=p^{\mu}\gamma_{\mu},\, L(\Lambda) \in O(3, 1),\,
L(-\Lambda) = L(\Lambda).
\label{LLambda}
\end{equation}

{\it Exercise 2.4} Verify that $\Lambda = exp(\lambda^{\mu\nu}\gamma_{\mu\nu})$,
where $(\lambda^{\mu\nu})$ is a skewsymmetric matrix of real numbers, satisfies
the last equation (\ref{beta}) and hence belongs to $Spin(3, 1)$. How does this
expression fit the one in Proposition 2.2 (b)? Prove that $\Lambda \in
\Gamma_{3,1}$ iff $N(\Lambda) \in {\mathbb R}^*$. Verify that $c^* \beta =\beta
c$ for $c = a + \omega b, c^* = a - \omega b$ and that $\Lambda^{-1} =
\Lambda^\dagger$.

The resulting (4-dimensional) representation of $Spin(3, 1)$ (unlike that of
$Spin(3, 2) \simeq Sp(4, {\mathbb R})$) is reducible and splits into two complex
 conjugate representations, distinguished by the eigenvalues ($\pm i$) of the
central element $\omega$ of $Cl^0(3, 1)$. These are the (left and right)
{\it Weyl spinors}.

{\it Remark 2.2} If we restrict attention to the class of representations for which the Clifford units are either hermitean or antihermitean then the (anti)hermitean units would be exactly those for which $\gamma_\mu^2 = 1 (-1)$. Within this class the matrix $\beta$, assumed hermitean, is determined up to a sign; we shall choose it as $\beta = i\gamma^0$. This class is only preserved by unitary similarity transformations. By contrast, the implicit definition of the notion of hermitean conjugation contained in Proposition 2.2 (a) is basis independent.

{\it Exercise 2.5} Prove that the Lie algebra $spin(4) \subset Cl^0(4)$ splits
into a direct sum of two $su(2)$ Lie algebras. The Coxeter element $\omega$ has
eigenvalues $\pm 1$ in this case and the idempotents $1/2(1 \pm \omega)$ project
on the two copies of $su(2)$ (each of which has a single 2-dimensional
irreducible representation).

{\it Remark 2.3} Denote by $cl(p, q)$ the maximal semisimple Lie algebra (under
 commutation) of $Cl(p, q),\, p + q = n$. The following list of identifications (whose verification is left to the reader) summarizes and completes the examples of this section:
\begin{eqnarray}
&& cl(2) = sl(2, {\mathbb R}) = cl(1, 1);           \nonumber  \\
&& cl(3) = spin(3, 1) \simeq sl(2,{\mathbb C})=cl(1,2),    \nonumber  \\
&& cl(2, 1) = spin(2, 1)\oplus spin(2,1)\simeq sl(2,{\mathbb R})\oplus sl(2,
{\mathbb R}); \nonumber   \\
&& cl(4) = spin(5, 1) \simeq sl(2, {\mathbb H}) = cl(1,3), \, \,
 cl(3, 1) = spin(3, 3) \simeq sl(4,{\mathbb R}); \nonumber   \\
&& cl(5) = spin(5, 1)\oplus spin(5,1)\simeq sl(2,{\mathbb H})\oplus sl(2,
{\mathbb H}); \nonumber \\
&& cl(4, 1) = sl(4,{\mathbb C}) = cl(2, 3); \, \,
cl(3,2)= sl(4,{\mathbb R})\oplus sl(4,{\mathbb R}); \nonumber \\
&&  cl(6) = su(6, 2) = cl(5, 1);  \nonumber \\
&& cl(7) = sl(8, {\mathbb C}), \, \, cl(6, 1) = sl(4,{\mathbb H}); \nonumber \\
&& cl(8) = sl(16, {\mathbb R}), \, \, cl(7, 1) = sl(8,{\mathbb H}); \nonumber \\
&& cl(9)\simeq cl^0(9, 1) = sl(16,{\mathbb R}) \oplus sl(16, {\mathbb R}) \,
cl(8,1)=sl(16, {\mathbb R}).
\label{cl}
\end{eqnarray}

Here is also a summary of low dimensional Spin groups ($Spin(p, q) \in Cl^0(p, q) $) (see \cite{D88}, Table 4.1):
\begin{eqnarray}
&& Spin(1,1)={\mathbb R}_{>0}, Spin(2)= U(1); \, Spin(2,1)=
SL(2,{\mathbb R}), \, Spin(3) = SU(2);                      \nonumber \\
&& Spin(2,2)= SL(2,{\mathbb R})\times SL(2,{\mathbb R}), \, Spin(3,1) = SL(2,{\mathbb C}), \, Spin(4) = SU(2)\times SU(2);  \nonumber \\
&& Spin(3,2)=Sp(4,{\mathbb R}), \, Spin(4,1)=Sp(1,1;{\mathbb H}), \,
Spin(5)= Sp(2,{\mathbb H});              \nonumber    \\
&& Spin(3,3)=SL(4,{\mathbb R}), \, Spin(4,2)=SU(2,2), \nonumber  \\
&& Spin(5,1)= SL(2,{\mathbb H}), \,   Spin(6)=SU(4).
\label{Spin}
\end{eqnarray}

\newpage

\section{The Dirac $\gamma$-matrices in euclidean and\\ in Minkowski space}
\setcounter{equation}{0}
\renewcommand\theequation{\thesection.\arabic{equation}}

We shall now turn to the familiar among physicists {\it matrix representation}
of the Clifford algebra and use it to characterize in an alternative way the
properties ${\rm mod} 8$ of $Cl(D)$ and $Cl(D-1, 1)$, the cases of main
interest. As we have seen (see Table 1) if $p-q \ne 1\ {\rm mod} \ 4$, in
particular, in all cases of physical interest in which the space-time dimension $D$ is even, $D = 2m$, there is a unique irreducible ($2^m$-dimensional) representation of the associated Clifford algebra \cite{R}. It follows that for such $D$ any two realizations of the $\gamma$-matrices are related by a similarity transformation (for $Cl(4)$ this is the content of the {\it Pauli \footnote{Wolfgang Pauli (1900-1958), Nobel Prize in Physics, 1945 (for his exclusion principle), predicted the existence of a neutrino (in a letter ``to Lise Meitner (1878-1968) et al.'' of 1930 - see \cite{P}); he published his lemma about Dirac's matrices in 1936} lemma}). We shall use the resulting freedom to display different realizations of the $\gamma$-matrices for $D = 4$, suitable for different purposes.

It turns out that one can represent the $\gamma$-matrices for any $D$ as tensor products of the $2\times 2$ Pauli $\sigma$-matrices \cite{P27} (cf. \cite{BW} \cite{D88}) in such a way that the first $2m$ generators of $Cl(2m+2)$ are obtained from those of $Cl(2m)$ by tensor multiplication (on the left) by, say, $\sigma_1$. The generators of $Cl(2m+1)$ give rise to a reducible subrepresentation of $Cl(2m+2)$ whose irreducible components can be read off the represetnation of $Cl(2m)$:
\begin{eqnarray}
&&Cl(1): \{\sigma_1\};\quad  Cl(2(3)):  \{\sigma_i,\ i = 1, 2, (3)\};\nonumber
\\ &&Cl(4): \{\gamma_i = \sigma_1 \otimes \sigma_i,\ i= 1, 2, 3,\
\gamma_4 = \sigma_2 \otimes 1 \ ;                           \nonumber   \\
&&Cl(6): \Gamma_\alpha = \sigma_1 \otimes \gamma_\alpha,\ \alpha = 1, ..., 5; \
Cl(8): \Gamma^{(8)}_a = \sigma_1 \otimes \Gamma_a,\ a = 1, ...,7;  \nonumber  \\
&&Cl(10): \Gamma^{(10)}_a = \sigma _1 \otimes \Gamma^{(8)}_a, a = 1, ..., 9, \, \,
 {\rm where} \, \Gamma^{(2m)}_{2m} = \sigma_2 \otimes 1^{\otimes (m-1)},\nonumber
\\ && \Gamma_{2m+1}^{(2m)} = \sigma_3 \otimes 1^{\otimes (m-1)} = i^{3-m}
\omega_{2m-1,1},
\label{Gamm8}
\end{eqnarray}
where $1^{\otimes k} = 1 \otimes ... \otimes 1$ (k factors), $1$ stands for the $2\times 2$ unit matrix. The Clifford algebra $Cl(D-1, 1)$ of D-dimensional Minkowski\footnote{Hermann Minkowski (1864-1909) introduced his 4-dimensional space-time in 1907, thus completing the special relativity theory of Lorentz, Poincar\'e and Einstein.} space is obtained by replacing $\gamma_D$ by
\begin{equation}
\gamma^0 = i\gamma_{2m}\ (= -\gamma_0) \, \rm for \, D=2m, 2m+1.
\label{E-M}
\end{equation}

Note that while $\Gamma_{2m+1}^{(2m)}$ is expressed in terms of a product of
$\Gamma_a^{(2m)}, a\leq 2m$, the element $\Gamma_{2m+1}^{(2m+1)}$ is an
independent
Clifford unit. In particular, we only know that the product $\omega_3$ of
$\sigma_i, i = 1, 2, 3$ in $Cl(3)$ is a central element of square $-1$, while
we have
the additional relation $\sigma_1 \sigma_2 = i \sigma_3$ in $Cl(2)$, in accord
with the fact that the real dimension $(8)$ of $Cl(3)$ is twice that of $Cl(2)$.
Furthermore, as one can read off Table 1, the Clifford algebra $Cl(5)$ (or, more
generally, $Cl(q+5, q)$) is reducible so that the matrices $\Gamma_a^{(2m)}$ in
(\ref{Gamm8}) for $1\leq a \leq 2m+1$ realize just one of the two irreducible
components. Furthermore,
$\gamma_{2m+1}$ is proportional to $\omega(p, q), p + q = 2m, q = 0, 1$ but
only belongs, for $q = 1$, to the complexification of $Cl(p, q)$; for instance,
\begin{equation}
\gamma_5 = i \omega(3, 1) (= \sigma_3 \otimes 1).
\label{gamma5}
\end{equation}
The algebra $Cl(4, 1)$, which contains $\gamma_5$ as a real element, plays an
important role in physical applications that seems to be generally ignored.
Its Coxeter element $\omega(4, 1)$, being central of square $-1$, gives rise to
 a complex structure (justifying the isomorphism $Cl(4, 1) = {\mathbb C}[4]$
that can be read off Table 1). The Lie algebra $cl(4, 1) = sl(4,{\mathbb C})$ (see
(\ref{cl})) has a real form $su(2, 2) =\{x \in cl(4, 1); x^\dagger = - x\}$;
the corresponding Lie group is the {\it spinorial conformal group} $SU(2, 2) =
\{\Lambda \in Cl(4, 1); \Lambda^\dagger = \Lambda^{-1}\}$ which preserves the
{\it pseudohermitean form} ${\tilde \psi} \psi$.

We proceed to defining {\it(charge) conjugation}, in both the Lorentzian and the
euclidean framework, and its interrelation with $\gamma_{2m+1}$ for $D = 2m$.
This will lead us to the concept of {\it $KO$-dimension} which provides another
mod 8 characteristic of the Clifford algebras. (It has been used in the
noncommutative geometry approach to the standard model (see \cite{ChC},\cite{CC}
\cite{Ca} for recent reviews and references and \cite{Bar} for the Lorentzian
case).

We define the {\it charge conjugation matrix} by the condition
\begin{equation}
-\gamma_a^t C = C \gamma_a
\label{C}
\end{equation}
which implies
\begin{equation}
-\gamma_{ab}^t C = C \gamma_{ab} \ (2 \gamma_{ab} = [\gamma_a, \gamma_b]),
\label{spin}
\end{equation}
but
\begin{equation}
\gamma_{abc}^t C = C \gamma_{abc} \ (6\gamma_{abc} = [\gamma_a,
[\gamma_b, \gamma_c]]_+ - \gamma_b \gamma_a \gamma_c + \gamma_c \gamma_a
\gamma_b = -6\gamma_{bac}= 6\gamma_{cab}= ...).
\label{3gam}
\end{equation}
(In view of (\ref{E-M}), if (\ref{C}) is satisfied in the euclidean case, for
$\alpha = 1, ..., D$, then it also holds in the Lorentzian case, for $\mu = 0,
 ..., D-1$.) It is straightforward to verify that given the representation
(\ref{Gamm8}) of the $\gamma$-matrices there is a unique, up to a sign, choice of the charge conjugation matrix $C(2m)$ (for an even dimensional space-time) as a product of $Cl(2m-1, 1)$ units:
\begin{eqnarray}
&& C(2) = c := i\sigma_2, C(4) = \gamma_3 \gamma_1 = 1 \otimes c,
C(6) = \Gamma^0 \Gamma_2 \Gamma_4 = c \otimes \sigma_3 \otimes c, \nonumber \\
&& C(8) = \Gamma_1 \Gamma_3 \Gamma_5 \Gamma_7 = 1 \otimes c \otimes \sigma_3
\otimes c, (\Gamma_a \equiv \Gamma^{(8)}_a),                   \nonumber \\
&& C(10) = \Gamma^0 \Gamma_2 \Gamma_4 \Gamma_6 \Gamma_8 = c \otimes \sigma_3
\otimes c \otimes \sigma_3 \otimes c.
\label{C2m}
\end{eqnarray}
The above expressions can be also used to write down the charge conjugation matrix for odd dimensional space times. A natural way to do it is to embed $Cl(2m-1)$ into $Cl(2m)$ thus obtaining a reducible representation of the odd Clifford algebras. Then we have two inequivlent solutions of (\ref{C}):
\begin{eqnarray}
 && C(2m-1):= C(2m) \Rightarrow {\overline C}(5) C(5)={\overline C}(7) C(7) = 1 =
- {\overline C}(3) C(3).           \nonumber             \\
&& C'(2m-1) = i^{5-m}\omega_{2m-1} C(2m) (= -i^{5-m} C(2m)\omega_{2m-1}), \nonumber \\
&& \Rightarrow {\overline C}'(2m-1)C'(2m-1) = -{\overline C}(2m-1)C(2m-1).
\label{Codd}
\end{eqnarray}
In particular, $C(5)$ and $C'(5)$ (satisfying (\ref{C}) for $1\leq a \leq 5$)
only exist in a reducible 8-dimensional representation of $Cl(4, 1) (\in
Cl(5, 1))$. (We observe that, with the above choice of phase factors, all
matrices $C$ are real.)

We define (in accord with \cite{ChC}) the {\it euclidean charge conjugation
operator} as an antiunitary operator $J$ in the $2^m$-dimensional complex
Hilbert space $\mathcal{H}$ (that is an irreducible Clifford module - i.e., the
 (spinor) representation space of $Cl(2m)$) expressed in terms of the matrix
$C(2m)$ followed by complex conjugation:
\begin{equation}
J = K C(2m) \Rightarrow J^2 = {\bar C}(2m) C(2m) = (-1)^{m(m+1)/2}.
\label{JC}
\end{equation}
We stress that Eq. (\ref{JC}) is independent of possible $i$-factors in $C$
(that would show up if one assumes that $C(2m)$ belongs, e.g., to $Cl(2m)$).

Alain Connes \cite{C06} defines the {\it KO dimension} of the (even dimensional)
 noncommutative internal space of his version of the standard model by two
signs: the sign of $J^2$ (\ref{JC}) and the factor $\epsilon(m)$ in the
commutation relation between $J = J(2m)$ and the {\it chirality operator}
$\gamma := \gamma_{2m+1}$:
\begin{equation}
J \gamma = \epsilon(m) \gamma J \, \, (\gamma = \gamma^*, \gamma^2 = 1).
\label{Jgam}
\end{equation}

Since $\gamma_{2m+1}$ of (\ref{Gamm8}) is real the second sign factor is
determined by the commutation relation between $C(2m)$ and $\gamma_{2m+1}$; one
 finds
\begin{equation}
\epsilon(m) = (-1)^m.
\label{eps}
\end{equation}
The signature, $(+, -)$, needed in the noncommutative geometry approach to the
standad model (see \cite{ChC}), yields KO dimension 6  mod  8 of the internal
space (the same as the dimension of the compact Calabi-Yau manifold appearing in
superstring theory).

The charge conjugation operator for Lorentzian spinors involves the matrix
$\beta$ of Eq. (\ref{beta}) that defines an invariant hermitean form in
${\mathbb C}^4$ (multiplied by an arbitray phase factor $\eta$ which we shall choose to make the matrix $\eta \beta C$ appearing in (\ref{JLB}) below real):
\begin{eqnarray}
&& J_L = K \eta \beta C \Rightarrow J_L^2 = {\bar B} B \, {\rm  where} \,
B:= \eta \beta C (= \gamma^0 C)  \nonumber  \\
&& \Rightarrow B^t = (= B^*) = B \, {\rm for} \, Cl(p,1), p = 1, 2, 3 \, \mathrm{mod} 8
\nonumber \\
&& B^t = - B \,(B^2 = -1) \, {\rm otherwise}.
\label{JLB}
\end{eqnarray}
It follows that $J_L^2$ has the opposite sign of $J^2$.
It is easy to verify that $\epsilon(m)$ also changes sign when using the charge conjugation for Lorentzian signature:
\begin{equation}
J_L^2 = - J^2, \, \epsilon_L(m) = - \epsilon(m).
\label{EM}
\end{equation}
In both cases the above two signs in a space-time of dimension $2m$ (and hence the KO dimension) is periodic in $m$ of period $4$.

Whenever $J^2 = 1$, the charge conjugation allows to define the notion of
{\it real} or {\it Majorana spinor}.  Indeed, in this case $J$ admits the
eigenvalue $1$ and we shall say that $\psi$ is a Majorana spinor if $J\psi =
\psi$. It is clear from Table 1 that Majorana spinors only exist for signatures
$p-q=0, 1, 2$ mod 8 ($p(=D-1)=1, 2, 3$ for $Cl(p,1)$ (\ref{JLB}).

{\it Exercise 3.1}  Prove that $J\Lambda J = \Lambda$ for $J^2 = 1, \Lambda \in Spin(p, q)$ so that the above reality property is $Spin(p, q)$-invariant.

Since the chirality operator (which only exists in dimension $D=2m$) has square $1$ (according to (\ref{Jgam})) it has two eigenspaces spanned by two $2^{m-1}$-dimensional {\it Weyl\footnote{Hermann Weyl (1885-1955) worked in G\"ottingen, Z\"urich and Princeton. He came as close as anyone of his generation to the universalism of Henri Poincar\'e and of his teacher David Hilbert (1862-1943). He introduced the 2-dimensional spinors in $Cl(3, 1)$ for a ``massless electron'' in \cite{W}; he wrote about spinors in $n$ dimensions in a joint paper with the German-American mathematician Richard Brauer (1901-1977) in 1935  \cite{BW}.} spinors}. They are complex conjugate to each other for $p-q = 2$ mod 4 (i.e. for $Cl(2), Cl(3, 1), Cl(6),
Cl(7, 1)$); self-conjugate for $p - q = 4$ (for $Cl(4), Cl(5, 1)$; they are
(equivalent to) real {\it Majorana-Weyl spinors} for $p-q = 0$ mod 8
(1-dimensional for $Cl(1, 1)$, 8-dimensional for $Cl(8)$, 16 dimensional for
$Cl(9, 1)$).

Consider the simplest example of a Majorana-Weyl field starting with the
massless
 \emph{Dirac equation} in the $Cl(1,1)$ module of 2-component spinor-valued
functions $\psi$ of $x=(x^0,x^1):$
\begin{eqnarray}\label{3.13}
    &\gamma\partial\psi\equiv(\gamma^0\partial_0+\gamma^1\partial_1)\psi=0,&\\
    &\gamma^0=c=\left(\!\!
                \begin{array}{cc}
                   \phantom{-}0 & 1 \\
                   -1 & 0 \\
                \end{array}
               \!\!\right),\,
    \gamma^1=\sigma_1\equiv\left(\!\!
                            \begin{array}{cc}
                              0 & 1 \\
                              1 & 0 \\
                            \end{array}
                          \!\!\right),\,
    \partial_{\nu}=\frac{\partial}{\partial x^{\nu}}&\nonumber
\end{eqnarray}
The chirality operator is diagonal in this basis,
 so that the two components of $\psi$ can be interpreted as "left and right":
\begin{equation}\label{Chiralcomp}
\gamma=\gamma^0\gamma^1=\sigma_3=\left(\!\!
                \begin{array}{cc}
                  1 & \phantom{-}0 \\
                   0 & -1 \\
               \end{array}
               \!\!\right)\;\Rightarrow\;
              \Psi=\left(\!\!
                               \begin{array}{c}
                                       \Psi_L \\
                                       \Psi_R \\
                               \end{array}
                         \!\!\right).
\end{equation}
Thus equation (\ref{3.13}) can be written as a (decoupled!)
system of \emph{Weyl equations}:
\begin{equation}\label{3.15}
(\partial_0+\partial_1)\psi_R=0=(\partial_1-\partial_0)\psi_L,
\end{equation}
implying that the chiral fields behave as a \emph{left} and \emph{right} movers:
\begin{equation}\label{left-right movers}
    \psi_L=\psi_L(x^0+x^1),\;\psi_R=\psi_R(x^0-x^1).
\end{equation}
A priori $\psi_c,\,c=L,R$ are complex valued functions, but since the
coefficients of the Dirac equation are real $\psi_c$ and $\bar{\psi}_c$
satisfy the same equation, in particular, they can be both real. These are the
(1-component) Majorana-Weyl fields (appearing e.g. in the chiral Ising model
- see for a review \cite{FST}).

{\it Exercise 3.2} Prove that there are no Majorana-Weyl solutions of the Dirac
 equation $(\sigma_1\partial_1+\sigma_2\partial_2)\Psi_E=0$ in the $Cl(2)$
module ( $E$ standing for Euclidean), but there is a 2-component Majorana
spinor  such that the two components of $\Psi_E$ are complex conjugate to each
other.

We are not touching here the notion of {\it pure spinor} which recently gained popularity in relation to (multidimensional) superstring theory - see \cite{BB} for a recent review and \cite{U} for a careful older work involving 4-fermion identities.

{\bf Historical note}. The enigmatic genius {\it Ettore Majorana} (1906-1938(?))
  has fascinated a number of authors. For a small sample of writngs about him -
 see (in order of appearance) \cite{P82}, \cite{GR}, \cite{Z06}, \cite{E08},
and Appendix A to \cite{B10} (where his biography by E. Amaldi in Majorana's
collected work is also cited). Let me quote at some length the first hand
impressions of Majorana of another member of the``circle of Fermi'', Bruno
Pontecorvo (for more about whom - see the historical note to Sect. 5): ``When I
 joined as a first year student the Physical Institute of the Royal University
of Rome (1931) Majorana, at the time 25 years old, was already quite famous
within the community of a few Italian physicists and foreign scientists who were
 spending some time in Rome to work under Fermi. The fame reflected first of all
 the deep respect and admiration for him  of Fermi, of whom I remeber exactly
these words: ''once a physical question has been posed, no man in the world is
capable of answering it better and faster than Majorana``. According to the
joking lexicon used in the Rome Laboratory, the physicists, pretending to be
associated with a religious order, nicknamed the infalliable Fermi as the Pope
and the intimidating Majorana as the Great Inquisitor. At seminars he was
usually silent but occasionally made sarcastic and paradoxical comments, always
 to the point. Majorana was permanently unhappy with himself (and not only with
 himself!). He was a pessimist, but had a very accute sense of humour. It is
difficult to imagine persons as different in character as Fermi and Majorana...
 Majorana was conditioned by complicated ... living rules ... In 1938 he
literally disappeared. He probably committed suicide but there is no absolute
certainty about this point. He was quite rich and I cannot avoid thinking that
his life might not have finished so tragically, should he have been obliged to
work for a living.'' Majorana thought about the neutron before James Chadwick
(1891-1974) discovered it (in 1932 and was awarded the Nobel Prize for it in
1935) and proposed the theory of ``exchange forces'' between
the proton and the neutron. Fermi liked the theory but Ettore was only convinced to publish it by Werner Heisenberg (1901-1976) who was just awarded the Nobel Prize in Physics when Majorana visited him in 1933.
Majorana was not happy with Dirac's hole theory of antiparticles (cf. the discussion in \cite{P10}). In 1932, in a paper ``Relativistic theory of particles with arbitrary intrinsic angular momentum'' (introducing the first infinite dimensional representation of the Lorentz group) he devised an infinite component wave equation with no antiparticles (but with a continuous tachyonic mass spectrum). His last paper \cite{M37} that was, in the words of \cite{P82}, forty years ahead of its time, is also triggered by this dissatisfaction\footnote{According to the words, which A. Zichichi \cite{Z06} ascribes to Pontecorvo, it was Fermi who, aware of Majorana's reluctance to write up what he has done, wrote himself the article, after Majorana explained his work to him.}. In its summary (first translated into English by Pontecorvo) he acknowledges that for electrons and positrons his approach may only lead to a formal progress. But, he concludes ``it is perfectly possible to construct in a very natural way a theory of neutral particles without negative (energy) states.'' The important physical consequence of the (possible) existence of a truly neutral (Majorana) particle - the {\it neutrinoless double beta decay} - was extracted only one year later, in 1938, by Wendel H. Furry (1907-1984) in what Pontecorvo calls ``a typical incubation paper ... stimulated by Majorana and (Giulio) Racah (1909-1965) thinking'' and still awaits its experimental test.

\bigskip

\section{Dirac, Weyl and Majorana spinors in\\ 4D  Minkowski space-time}
\setcounter{equation}{0}
\renewcommand\theequation{\thesection.\arabic{equation}}

For a consistent physical interpretation of spinors, one needs local anticommuting
(spin 1/2) quantum fields. (Their ''classical limit'' will produce an object which is unknown in physics: strictly anticommuting Grassmann valued fields.)We choose to build up the complete picture step by step, following roughly, the historical development.

\par To begin with, the Dirac spinors form a spinor bundle over 4-dimensional
space-time with a ${\mathbb{C}}^4$ fibre.(We speak of elements of a fibre
bundle, rather than functions on Minkowski space, since $\psi(x)$ is double
valued: it changes sign under rotation by $2\pi$.) The spinors span an
irreducible representation (IR) of $Cl(3,1)$ which remains irreducible when
restricted to the group $Pin(3,1)$, but splits into two inequivalent IRs of its
 connected subgroup $Spin(3,1)\simeq SL(2,\mathbb{C})$. These IRs are spanned by
 the 2-component ''left and right'' Weyl spinors, eigenvectors of the chirality
\begin{equation}\label{4.1}
   ( \gamma=)\gamma_5 = i\omega_{3,1}=i\gamma_0\gamma_1\gamma_2\gamma_3=\sigma_3
\otimes 1 =
    \left(\!\!
      \begin{array}{cc}
        1 &\phantom{-}0 \\
        0& -1\\
      \end{array}
    \!\!\right).
\end{equation}

{\it Remark 4.1} Relativistic local fields transform under finite dimensional
representations of $SL(2,{\mathbb C})$, the {\it quantum mechanical Lorentz
group} - see Sect. 5.6 of \cite{Wei} for a description of these representations
 targeted at applications to the theory of quantum fields. Here we just
note that the finite dimensional irreducible representations (IRs) of
$SL(2,{\mathbb C})$ are
labeled by a pair of half-integer numbers $(j_1, j_2), j_i = 0, 1/2, 1, ...$.
Each IR is spanned by {\it spin-tensors} $\Phi_{A_1...A_{2j_1}\dot{B}_1...
\dot{B}_{2j_2}}, A, \dot{B} = 1, 2$, symmetric with respect to the dotted and
undotted indices, separately; thus the dimension of such an IR is
$dim(j_1, j_2) = (2j_1+1)(2j_2+1)$. The Weyl
spinors $\psi_L$ and $\psi_R$, introduced below, transform under the basic
(smallest nontrivial) IRs $(1/2, 0)$ and $(0, 1/2)$ of $SL(2,{\mathbb C})$,
respectively. Their direct sum span the 4-dimensional Dirac spinors which
transform under an IR of $Pin(3, 1)$ (space reflection exchanging the two
chiral spinors).

The "achingly beautiful" (in the words of Frank Wilczek, cited in \cite{F}, p.142) Dirac equation
\cite{D28} for a free particle of mass $m$, carved on Dirac's commemorative stone in Westminster
Abbey, has the form\footnote{The first application of the Dirac equation dealt
with the fine structure of the energy spectrum of hydrogen-like atoms (see, e.g.
 \cite{Wei}, Sect. 14.1, as well as the text by Donkov and Mateev \cite{DM}).
It was solved exactly in this case by Walter Gordon in Hamburg and by Charles
Galton Darwin in Edinburgh, weeks after the appearance Dirac's paper (for a
historical account see \cite{MR}). Our conventions for the Dirac equation, the
chirality matrix $\gamma_5$ etc. coincide with Weinberg's text (see \cite{Wei}
Sect. 7.5) which also adopts the space-like Lorentz metric (but not with the
inscription at the Westminster Abbey).}
\begin{equation}\label{4.2}
    (m+\gamma\partial)\psi=0,\;\gamma\partial=\gamma^{\mu}\partial_{\mu},\;
    \psi=\left(\!\!
           \begin{array}{c}
            \psi_L \\
             \psi_R\\
           \end{array}
         \!\!\right),              \nonumber     \\
\widetilde{\psi} (m-\gamma\partial) = 0 \, {\rm for} \, \widetilde{\psi}
= \psi^*\beta
\end{equation}
(the partial derivatives in the equation for $\widetilde{\psi}$ acting to the
left).
Using the realization (\ref{Gamm8}) of the $\gamma$-matrices,
\begin{equation}\label{gammaMatrices}
    \gamma^0=\left(\!\!
      \begin{array}{cc}
        \phantom{-}0 & 1\\
        -1 & 0 \\
      \end{array}
    \!\!\right),\;
    \gamma^j=\left(\!\!
      \begin{array}{cc}
        0 & \sigma_j\\
        \sigma_j & 0 \\
      \end{array}
    \!\!\right),\; j=1,2,3,
\end{equation}
(where each entry stands for a $2\times 2$ matrix) we obtain the system of
equations
\begin{equation}\label{4.4}
    (\partial_0+\boldsymbol{\sigma\partial})\psi_R + m\psi_L = 0 =
    (\boldsymbol{\sigma\partial} -{\partial_0)\psi_L + m\psi_R.}
\end{equation}
(They split into two decoupled Weyl equations in the zero
mass limit, the 4-dimensional counterpart of (\ref{3.15}).)

We define the charge conjugate 4-component spinor $\psi^C$ in accord with
(\ref{JLB}) by
\begin{equation}\label{4.5}
    \psi^C=\psi^{*}\gamma^0 C, \; C=\left(\!\!
                                                  \begin{array}{cc}
                                                    c & 0 \\
                                                    0 & c \\
                                                  \end{array}
                                                \!\!\right)
(\Rightarrow(\psi^C)^C = \psi).
\end{equation}
One finds
\begin{equation}\label{4.6}
    B:=\gamma^0 C=\left(\!\!
                  \begin{array}{cc}
                    \phantom{-}0& c\\
                    -c & 0 \\
                  \end{array}
                \!\!\right) = B^t,\;
\psi_L^C=-\psi_R^{*}c,\;\psi_R^C=\psi_L^{*}c
\end{equation}
It was Majorana \cite{M37} who discovered (nine years after Dirac wrote his
equation) that there exists a real representation, spanned by $\psi_M$ of
$Cl(3,1)$, for which
\begin{equation}\label{4.7}
    \psi_M^C=\psi_M\;\Leftrightarrow\psi_R=\psi_L^{*}c\;(\psi_L=\psi_R^{*}c^{-1}).
\end{equation}
(Dirac equation was designed to describe the electron - a charged particle,
different from its antiparticle. Majorana thought of applying his ``real
spinors'' for the description of the neutrino, then only a hypothetical neutral
 particle - predicted in a letter by Pauli and named by Fermi \footnote{Enrico
Fermi (1901-1954), Nobel Prize in Physics, 1938, for his work
on induced radioactivity. It was he who coined the term {\it neutrino} - as a diminutive of neutron. (See E. Segr\`e, {\it Enrico Fermi - Physicist}, Univ. Chicago Press, 1970) }.)

Sometimes, the Majorana representation is defined to be one with real
$\gamma$-matrices. This is easy to realize (albeit not necessary) by just
setting $\gamma_2^M = \gamma_5$ (which will give $\gamma_5^M := i\gamma_0^M
\gamma_1^M \gamma_2^M \gamma_3^M = -\gamma_2$). In accord with Pauli lemma there
 is a similarity transformation (that belongs to $Spin(4, 1) \subset Cl^0(4, 1)
\simeq Cl(4)$) between $\gamma_\mu^M$ and $\gamma_\mu$ (of Eq.
(\ref{gammaMatrices})):
\begin{equation}
\gamma_\mu^M = S \gamma_\mu S^* \, {\rm for} \, S = \frac{1}{\sqrt 2}(1 -
\gamma_2 \gamma_5)  \, (S^* = \frac{1}{\sqrt 2}(1 + \gamma_2 \gamma_5)).
\label{gamM}
\end{equation}
The charge conjugation matrix $C_M$ in the Majorana basis coincides with
$\gamma_0^M$, the only skew-symmetric Majorana matrix while the symmetric form
$B_M$ of Eq. (\ref{JLB}) is $1$:
\begin{equation}
C_M = \gamma_0^M, \ B_M = \gamma_M^0 C_M = 1 \Rightarrow \psi^C = \psi^*.
\label{CM}
\end{equation}
We prefer to work in the {\it chirality basis} (\ref{4.1}) (called {\it Weyl
basis} in \cite{M84}) in which $\gamma_5$ is diagonal (and the Lorentz
/$Spin(3, 1)$-/ transformations are reduced).

{\it Exercise 4.1} Find the similarity transformation which relates the {\it
Dirac basis} (with a diagonal $\gamma_{Dir}^0$),
\begin{equation}
i\gamma_{Dir}^0 = \gamma_5, \gamma_{Dir}^j = \gamma^j \, \Rightarrow C_{Dir} = i
\gamma_2,
\label{gamD}
\end{equation}
to our chirality basis. Compute $\gamma_{Dir}^5$.

The Dirac quantum field $\psi$ and its conjugate
$\tilde{\psi}$, which describe the free electron and positron, are operator
valued solutions of Eq. (\ref{4.2}) that are expressed as follows in terms of
their Fourier (momentum space) modes:
\begin{eqnarray}\label{4.11}
 & \psi(x)=\int(a_{\zeta}(p)e^{ipx} u_{\zeta}(p)+b_{\zeta}^*(p) e^{-ipx}v_{\zeta}(p))(dp)_m&\nonumber \\
 &\widetilde{\psi}(x)(= \psi^*(x)\beta) = \int(a^*_{\zeta}(p)e^{-ipx}\widetilde{u}_{\zeta}(p)+b_{\zeta}(p) e^{ipx}\widetilde{v}_{\zeta}(p))(dp)_m,&
\end{eqnarray}
where summation in $\zeta$ (typically, a \emph{spin projection}) is understood,
spread over the two independent (classical) solutions of the linear homogeneous
(algebraic) equations
\begin{eqnarray}\label{4.12}
(m+ip\gamma)u_{\zeta}(p)\!\!&\!\!=\!&\!\!0\Big(=\widetilde{u}_{\zeta}(p)
(m+ip\gamma)\Big),\nonumber\\
(m-ip\gamma)v_{\zeta}(p)\!\!&\!=\!&\!\!0,\; {\mathrm{for}} \;
p^0=\sqrt{m^2+\boldsymbol{p}^2},
\end{eqnarray}
while $(dp)_m$ is the normalized Lorentz invariant volume element on the
positive mass hyperboloid,
\begin{equation}\label{4.13}
(dp)_m=(2\pi)^{-3}\frac{d^3p}{2p^0}=
\Big(\int^{\infty}_0\!\!\delta(m^2+p^2)dp^0\Big)
\frac{d^3p}{(2\pi)^3},\;p^2=\boldsymbol{p}^2-p_0^2.
\end{equation}
The \emph{creation} $(a^*_{\zeta},\,b^*_{\zeta})$ and the \emph{annihilation}
$(a_{\zeta},\,b_{\zeta})$ \emph{operators} are assumed to satisfy the {\it
covariant} canonical anticommutation relations

\begin{eqnarray}\label{4.14}
&&[a_{\zeta}(p),a^*_{\zeta'}(p')]_+=\delta_{\zeta\zeta'}(2\pi)^3 2p^0\delta(\textbf{p}-\textbf{p}')=[b_{\zeta}(p),b^*_{\zeta'}(p')]_+\nonumber\\
&&[a_{\zeta}(p),b^*_{\zeta'}(p')]_+= 0 = [a_\zeta, b_{\zeta'}]_+ =... .
\end{eqnarray}
\emph{Stability of the ground state} (or the \emph{energy positivity}) requires that the vacuum vector $|0\big>$ is annihilated by
$a_{\zeta},\,b_{\zeta}$:
\begin{equation}\label{4.15}
a_{\zeta}(p)|0\big>=0=\big<0|a^*_{\zeta}(p),\;
b_{\zeta}(p)|0\big>=0=\big<0|b^*_{\zeta}(p).
\end{equation}
This allows to compute the electron 2-point function
\begin{equation}\label{4.16}
\big<0|\psi(x_1)\otimes\widetilde{\psi}(x_2)|0\big>=
\int e^{ipx_{12}}(m-i\gamma p)(dp)_m,\;x_{12}=x_1-x_2,
\end{equation}
where we have fixed on the way the normalization of the solutions of Eq.
(\ref{4.12}),
\begin{eqnarray}\label{4.17}
&& \sum_\zeta u_\zeta(p)\otimes\widetilde{u}_\zeta(p) = m - i\gamma p, \,
\sum_\zeta v_\zeta(p)\otimes\widetilde{v}_\zeta(p) = -m -i\gamma p; \nonumber \\
&& \widetilde{u}_\eta(p) u_\zeta(p) = 2m \delta_{\eta \zeta} =
- \widetilde{v}_\eta(p) v_\zeta(p).
\end{eqnarray}

{\it Remark 4.2} Instead of giving a basis of two independent solutions of Eq.
(\ref{4.12}) we provide {\it covariant} (in the sense of (\ref{LLambda})) {\it
pseudohermitean expressions} for the sesquilinear combinations (\ref{4.17}).
The idea of using bilinear characterizations of spinors is exploited
systematically in \cite{L}.

Note that while the left hand side of (\ref{4.17}) involves (implicitly) the
matrix $\beta$, entering the {\it Dirac conjugation}
\begin{equation}\label{4.18}
u \rightarrow \widetilde{u} = \bar{u} \beta \, (\psi \rightarrow \widetilde
{\psi} = \psi^*\beta),
\end{equation}
its right hand side is independent of $\beta$; thus Eq. (\ref{4.17}) can serve
to determine the phase factor in $\beta$. In particular, it tells us that
$\beta$ should be hermitean:
\begin{eqnarray}
&& (\sum_\zeta u_\zeta(p)\otimes\widetilde{u}_\zeta(p))^* \beta = \beta^*
\sum_\zeta u_\zeta(p)\otimes\widetilde{u}_\zeta(p), \nonumber \\
&& (m - i \gamma p)^* \beta = \beta (m - i\gamma p) \, \Rightarrow \beta^* =
\beta = \beta^\dagger.
\end{eqnarray}
The positivity of matrices like $\sum_\zeta u_L(p,\zeta)\otimes \bar{u}_L(p,
\zeta)$ for the chiral components of $u$ (and similarly for $v$) - setting, in
particular, $\widetilde{u} = i(-\bar{u}_R, \bar{u}_L)$
 - fixes the remaining sign ambiguity (as $p^0>0$ according to (\ref{4.12}):
\begin{eqnarray}\label{4.20}
&& \beta = i\gamma^0 \Rightarrow \sum_\zeta u_L(p, \zeta)\otimes
\bar{u}_L(p,\zeta) = p^0 - {\bf p\sigma} =: \tilde{p}, \nonumber  \\
&& \sum_\zeta u_R(p,\zeta)\otimes\bar{u}_R(p,\zeta)= p^0 + {\bf p\sigma} =
\undertilde{p}
\, \, (\undertilde{p} \, \tilde{p} = -p^2 = m^2).
\end{eqnarray}
Using further the Dirac equation (\ref{4.12}),
\begin{equation}
m u_L = i\tilde{p} u_R, \, m u_R = -i \undertilde{p} u_L,
\end{equation}
we also find
\begin{equation}
\sum_\zeta u_L(p, \zeta) \otimes \bar{u}_R(p,\zeta) = im
= - \sum_\zeta u_R(p, \zeta) \otimes \bar{u}_L(p,\zeta).
\end{equation}

{\it Exercise 4.2} Deduce from (\ref{4.17}) and from the definition (\ref{4.5})
 of charge conjugation that $u^C$ can be identified with $v$; more precisely,
\begin{equation}\label{4.23}
\sum_\zeta u_\zeta^C(p)\otimes\widetilde{u^C}_\zeta(p) = -m - i\gamma p (=
\sum_\zeta v_\zeta(p)\otimes\widetilde{v}_\zeta(p)).
\end{equation}

Any Dirac field can be split into a real and an imaginary part (with respect to charge conjugation):
\begin{eqnarray}
&& \psi(x) = \frac{1}{\sqrt{2}} (\psi_M(x) +\psi_A(x)), \, \psi_M^C = \psi_M, \,
\psi_A^C = - \psi_A,            \nonumber            \\
&& \psi_M(x) =\int (c(p)u(p)e^{ipx} + c^*(p) u^C(p)e^{-ipx}) (dp)_m, \nonumber \\
&& \psi_A(x) = \int(d(p)u(p)e^{ipx} - d^*(p) u^C(p)e^{-ipx}) (dp)_m.
\end{eqnarray}
The field $\psi$ can then again be written in the form (\ref{4.11}) with
\begin{equation}
\sqrt{2} a(p) = c(p) + d(p), \, \sqrt{2} b(p) = c(p) - d(p).
\end{equation}

{\it Remark 4.3} For a time-like signature, the counterpart of the Majorana
representation for $Cl(1,3)$ would involve pure imaginary $\gamma$-matrices; the
free Dirac equation then takes the form $(i\gamma \partial - m)\psi = 0$
(instead of (\ref{4.2}). Such a choice seems rather awkard (to say the least)
for studying real spinors.

\bigskip

\section{Peculiarities of a Majorana mass term.\\ Physical implications}
\setcounter{equation}{0}
\renewcommand\theequation{\thesection.\arabic{equation}}

The Lagrangian density for the free Dirac field has the form
\begin{equation}\label{5.1}
{\mathcal{L}}_0 = -\widetilde{\psi}(m+ \gamma \partial) \psi.
\end{equation}
(In the quantum case one should introduce normal ordering of the fields, but
such a modification would not affect the conclusion of our formal discussion.)
The mass term, $m \widetilde{\psi} \psi$ is non-vanishing for a Dirac field
at both the quantum and the classical level (viewing, in the latter case, the
components of $\psi$ as /commuting/ complex-valued functions). This is, however,
 not the case for a Majorana field, satisfying (\ref{4.7}). Indeed, the
implication
\begin{equation}\label{5.2}
\psi^C = \psi \Rightarrow \widetilde{\psi} \psi = i\psi C^{-1} \psi
\end{equation}
of the reality of $\psi$ tells us that the mass term vanish for a classical
Majorana field since the charge conjugation matrix is antisymmetric (in four
dimensions). This is made manifest  if we insert, using (\ref{4.7}), the chiral
 components of the Majorana spinor:
\begin{equation}\label{5.3}
\widetilde{\psi} \psi = i(\psi_L^* \psi_R - \psi_R^* \psi_L) =
i(\psi_R c^{-1} \psi_R - \psi_L c \psi_L) \, (c = i\sigma_2 = - c^t).
\end{equation}

Thus, the first peculiarity of a Majorana mass term is that it would be a
purely quantum effect with no classical counterpart, in contrast to a naive
understanding of the ``correspondence principle''. An even more drastic
departure from the conventional wisdom is displayed by the fact that the
reality condition (\ref{4.7}) (or, equivalently, (\ref{5.2})) is not invariant
under phase transformation ($\psi \rightarrow e^{i\alpha} \psi$). Accordingly,
the $U(1)$ current of an {\it anticommuting}  Majorana  field,
\begin{equation}
i\widetilde{\psi}\gamma^\mu \psi = \psi C\gamma^\mu \psi
\end{equation}
vanishes since the matrix $C\gamma^\mu$ is symmetric as a consequence of the
definition of $C$, (\ref{C}). In particular, a Majorana neutrino coincides with
 its antiparticle implying a violation of the lepton number conservation, a
consequence that may be detected in a neutrinoless double beta decay (see
\cite{BP}, \cite{B10} \cite{R11} and references therin) and may be also in a
process of left-right symmetry restoration that can be probed at the Large
Hadron Collider (\cite{TV}, \cite{St}).

The discovery of neutrino oscillations is a strong indication of the existence
of positive neutrino masses (for a recent review by a living classic of the
theory and for further references - see \cite{B10}). The most popular theory of
neutrino masses, involving a mixture of Majorana and Dirac neutrinos, is based
on the so called {\it ``seesaw mechanism''}, which we proceed to sketch (cf.
\cite{K09} for a recent review with an eye towards applications to cosmological
dark matter and containing a bibliography of 275 entries).

A model referred to as $\nu$MSM (for \emph{minimal standard model with neutrino
 masses}) involves three Majorana neutrinos $N_a\; (a=1,2,3)$ on top of the
three known weakly interacting neutrinos, $\nu_{\alpha}$,  that are part of three
 leptonic lefthanded doublets $L_{\alpha}\,(\alpha=e,\mu,\tau)$. To underscore
the fact that $N_a$ are \emph{sterile neutrinos} which do not take part in the
standard electroweak interactions, we express them, using (\ref{4.7}), in terms of
{\it right handed} (2-component, Weyl) spinors $R_a$ and their conjugate,
\begin{equation}\label{5.5}
   N_a=\left(\!
         \begin{array}{c}
           R^*_a c^{-1} \\
           R_a\phantom{ c^{-1}}\\
         \end{array}
       \!\right),\; a=1,2,3.
\end{equation}

The $\nu$MSM action density is obtained by adding to the standard model
Lagrangian, ${\mathcal{L}}_{SM}$, an Yukawa interaction term  involving the
Higgs doublet $H$ along with $L_{\alpha}$ and $N_a$ and the free Lagrangian for
the heavy Majorana fields:
\begin{equation}\label{5.6}
{\mathcal{L}}={\mathcal{L}}_{SM}-{\widetilde{N}}_a(\gamma\partial+M_a)N_a-y_{\alpha a}(H^*\widetilde{L}_{\alpha}N_a+H\widetilde{N}_aL_{\alpha})
\end{equation}
with the assumption that
\begin{equation}\label{5.7}
|y_{\alpha a}\big<H\big>|\ll M_a,
\end{equation}
where $\big<H\big>$ is the vacuum expectation value of the Higgs field
responsible for the spontaneous symmetry breaking
that yields positive masses in the standard model.

In order to display the idea of the seesaw mechanism\footnote{This idea has
been developed by a number of authors starting with P. Minkowski, 1977 - see
for historical references \cite{K09} and for some new developments \cite{TV} \cite{S}.
Its counterpart in the noncommutative geometry approach to the standard model that
uses the euclidean picture (in which there are no Majorana spinors) is discussed in
\cite{St} \cite{JKSS}.} we consider a
two-by-two block of the six-by-six ``mass matrix''
\begin{equation}\label{5.8}
 \left(\!
   \begin{array}{cc}
     0& y\big<H\big> \\
     y\big<H\big>& M \\
   \end{array}
 \!\right).
\end{equation}

 It has two eigenvalues $M_N$ and $-m_\ell$ where under the assumption
(\ref{5.7}), $m_\ell << M_N$. Identifying $m_\ell$ with the light left neutrino
 mass, and $M_N$ with the mass of the heavy sterile neutrino we find,
approximately,
\begin{equation}
m_\ell \simeq \frac{(y<H>)^2}{M}, \, \, M_N \simeq M.
\end{equation}

{\bf Historical note}: {\it Bruno Pontecorvo}.  Wolfgang Pauli (see footnote 13), who predicted the neutrino
in a letter not destined for publication, did not believe that it could ever be observed. A physicist who did
believe in the experimental  study of the neutrinos was Bruno Pontecorvo\footnote{The ``Recollections and
reflections about Bruno Pontecorvo'' by S.S. Gershtein, available electornically in both the original Russian
and in English, give some idea of this remarkable personality - which also emerges in Pontecorvo's own
recollections \cite{P82}.} (1913-1993), aptly called Mr. Neutrino by his long-time (younger) collaborator Samoil
M. Bilenky (see \cite{B06}). He proposed (in a 1946 report) a method for detecting (anti)neutrino in nuclear
reactors, a methodology used by Frederick Reines (1918-1999) and Clyde Cowan (1919-1974) in their 1956 experiment
that led to the discovery of neutrino (for which the then nearly 80-year-old Reines shared the Nobel Prize in
Physics in 1995). Pontecorvo predicted that the muon neutrino may be different from the electron one and proposed
an experimental method to prove that in 1959. His method was successfully applied three years later in the
Brookhaven experiment for which J. Steinberger, L. Lederman and M.  Schwarz were awarded the Nobel Prize in 1988.
He came to the idea of neutrino oscillation in 1957 and from then on this was his favourite subject. Vladimir
Gribov (1930-1997) and Pontecorvo considered in 1969 the possibility of lepton number violation through a Majorana
mass term and applied their theory to the solar neutrino problem. Bilenky and Pontecorvo introduced the general
Majorana-Dirac mass term that is used in the seesaw mechanism \cite{BP78}. (See for details and references
\cite{B06}.) Neutrino oscillations are now well established in a number of experiments - and await another Nobel Prize triggered by the formidable intuition of Bruno Pontecorvo.

{\bf Acknowledgments}. Clifford algebras have fascinated mathematical physicists all over the world. I have benefited, in particular, from conversations with Petko Nikolov and Ludwik Dabrowski who have popularized them at the University of Sofia and in Italy, respectively (see \cite{NY} and \cite{D88}). I also thank
Samoil Mihelevich Bilenky and Serguey Petcov for teaching me the physics of
Majorana neutrinos.

The author thanks the High Energy Division of The Abdus Salam International
Centre for Theoretical Physics (ICTP) and the Theory Group of the Physics
Department of CERN where these notes were completed. Partial support by grant
DO 02-257 of the Bulgarian National Science Foundation is gratefully
acknowledged.

\bigskip


\end{document}